\documentclass{article}

\usepackage{amssymb}

\usepackage{epsfig}

\usepackage{lscape}

\def\simpropto{\lower.2ex\hbox{$\; \buildrel \propto \over \sim \;$}}
\def\ltsim{\lower.5ex\hbox{$\; \buildrel < \over \sim \;$}}
\def\gtsim{\lower.5ex\hbox{$\; \buildrel > \over \sim \;$}}

\usepackage{graphicx,color}

\usepackage{graphicx,psfrag,amsmath}

\usepackage{yfonts}

\begin{document}

\title{SWAMPLAND REVISITED}
\author{ 
Michel Cass\' e \\
Directeur of Recherche (retired)  CEA, Saclay, F-91191, Gif-sur-Yvette, France
\\  
Joseph Silk\\
Department of Physics and Astronomy, The Johns Hopkins University,  \\ Baltimore MD 21218, USA\\
Institut d'Astrophysique de Paris, UMR7095:CNRS 
 UPMC-Sorbonne University, \\ F-75014, Paris, France\\
 Beecroft ~Institute~ of~ Particle~ Astrophysics~ and Cosmology, Department~ of~ Physics\\
 University~ of~ Oxford, Oxford OX1 3RH, UK\\
}

\maketitle

\begin{abstract}  
The transcendental expectation of string theory is that the nature of the fundamental forces, particle spectra and masses, together with coupling constants, is uniquely determined by mathematical and logical consistency, non-empirically, that is by pure reason.  However   pluralism triumphed with the explosive emergence of the multiverse. String theorists have extended a long-sought dream (their unique and final theory) to a landscape or a happy caparnaum. Proponents of string theory try to qualify their arguments via  swampland conjectures while cosmologists retreat to their telescopes. We review the current status of the string theory swampland.
\end{abstract}

We live near a star called the Sun in a republic of stars called the Milky Way, one of the billions of galaxies composing the Universe, where the vacuum energy is positive. This should not be the case according to the superstring ansatz embodied by Cumrum Vafa and others
\cite{2005hep.th....9212V,
2019ForPh..6700037P,
2021Univ....7..273G,
2021arXiv210201111V}.
Why? Lets begin with basics. We will critically review the swampland conjecture for the benefit of our astrophysics colleagues

\section {Swampland conjectures: out of the mud}
Instead of modeling particles as zero-dimensional points, string theory envisions them as tiny twigs undergoing diverse modes of vibrations corresponding to mass and charge.
The String Theory {\it landscape} or abstract panorama exhibits a huge number of semiclassical field theories.  The numbers of possible theories have exploded. The problem is how to find, from this huge number, the most suitable theory, if any,  that is consistent with our world.
To anticipate our conclusion, this vision that   {\it seems consistent but possibly not} illustrates the fuzziness of the concepts held by landscape and swampland advocates. In the following, for the sake of clarity, we split the problem into digestible slices.
Firstly, what is the Swampland? Its more of a program than a geographical topology. The Swampland program aims at distinguishing effective theories which can be extended into the realm of quantum gravity at high energy (in the ultraviolet) from those which cannot. 

Secondly, what is a semi-classical theory? In this restricted category, advocated by Stephen Hawking, matter is quantified, but not spacetime. In this uncertain, non-empirical  and debatably oxymoric framework, as emphasized elsewhere \cite{2014Natur.516..321E,2016arXiv160901966R} the impossible occurs:  black holes shine. This is one of the most  profound predictions in physics, hitherto unverified.

Next, what is an effective theory? We all work in terms of effective theories. We find descriptions that match what we actually see, interact with, and measure. Newton's laws are approximations that work at relatively low velocities and for large macroscopic objects. 
The combination of General Relativity and Quantum Field Theory along with the $\Lambda$CDM cosmological model are effective theories or approximations of quantum gravitation and quantum cosmology, respectively.

A good effective theory should tells us its limitations--the conditions and values of parameters for which the theory breaks down. This notion is practical and valuable. It is valuable up to a certain point or more exactly to a certain energy, a kind of Heisenberg cut but in cosmology. The laws of the effective theory succeed until we reach its limits when these assumptions are no longer true or our measurements or requirements become increasingly precise.

What lies beyond might be a more fundamental truth. And beyond, or within, lies an even more fundamental truth. Then, what indeed  is fundamental? This is a crucial ssue but would be too long to discuss in this pedagogical perspective, where we adopt a reductionist stance rather than an holistic/collective one.

As a quantum description, string theory  has the potential to incorporate gravity, at the expense, however, of increasing the number of space dimensions. In its extended version,  11-dimensional M theory is still a vibrant area of research but it  lacks  experimental or observational support and therefore it searches for justification in cosmological observations. Here  success is not guaranteed, on the contrary. So what  is right? String theory or cosmology?

The greatest difficulty with this type of theory is being able to test it with observations. Then one could circumvent the problem, or at least avoid it,  by studying the effective field theories which are semi-classical approximations. These theories can be tested experimentally, at least up to energy regimes that are so extreme as to require passage to a real String Theory. 
The Swampland Conjecture is based on the assumption of  finding general characteristics that a semi-classical effective field theory must satisfy to be considered a truly coherent theory. These are restrictions on the finiteness of the scalar field  or lower limits that the derivatives of the potentials present in these theories must satisfy.

\subsection {First swampland criterion}
This first criterion serves to select those models at very small distance, (extreme energies) compatible with early cosmological times, close to the Big Bang, which must be stable and converging near the singularity. We are obviously dealing with effective field theories in which the variation in absolute module of the field itself must not exceed a certain value  which in Planck units is of the order of unity.

\subsection {Second swampland  criterion}
This serves to select models suitable for describing  late times, our time, and it is a restriction on the slope of the potential with respect to the value of the potential itself. The criterion purports that this scenario does not derive from the presence of a positive cosmological constant, but from the existence of a potential dependent on a scalar field that is rolling along its profile, such as the inflationary potentials that we will consider in our analysis. Imposing that  $\left\lvert V \Delta V \right\rvert $ when $V > 0$ has a lower bound of the order of 1 in Planck units means that the potential has a small slope when the contribution of Dark Energy is equally small.

\subsection {de Sitter denial}
A positive vacuum energy can be realized by a scalar field potential with a local minimum leading to a long-lived de Sitter universe with an entropy proportional to the surface of the cosmological horizon. However it could be that the potential is positive but the scalar field does not rest at a minimum. This is the case in quintessence models, as long as the absolute value of the gradient of the potential  is small and of the order of the potential itself \cite{2018arXiv180608362O}.  Such an option is adopted by C.Vafa and others, to confront the difficulty in producing metastable de Sitter vacua in string theory. One consequence is that  dS space, the attached cosmological constant $\Lambda$, and the standard model of cosmology $\Lambda$CDM, all sink  in the swampland of failed theories.

Given this extraordinary claim, it is legitimate to ask if the condemnation of dS is definitive and irrevocable. The main argument for anti-$\Lambda$ is probably the strangeness of dS, which is maximally symmetric but not supersymmetric \cite{2018IJMPD..2730007D}. But one must admit that it is not strictly possible to construct SUSY theories in de Sitter space. Formally, dS superalgebra cannot be justified unless the action is constructed with matter coupled to the wrong sign of the kinetic  term.  KKLT overcome this objection  by uplifting AdS vacua into dS ones.  and the proposal that there might be no dS in superstring theory is fervently debated \cite{2019JHEP...04..095D, 2019JHEP...11..075G, 2021arXiv211210779C}.

\section {String tourism and ecology}
The eloquence of the String Theory Landscape masks the confrontation of the practitioners (not necessarily believers) before the proliferation of false vacua. The contents are not necessarily strings, but more general higher dimensional dynamical objects called branes. The formalism is  not yet a theory, rather a paradigm or a fashion, but rich in universes \cite{2003dmci.confE..26S}. When we refer to a de Sitter vacuum of string theory, we mean a vacuum which is a local minimum of the scalar potential where the potential takes a positive value. 
The choices for reconciliation of the huge initial value of the cosmological constant required by inflation range from an abrupt phase transition to a slowly rolling quintessence field.

The bulk of observations is consistent with a dark energy component of current energy density $10^{-120} M_{Pl}$  where $M_{Pl} = 1/ \sqrt {8\pi G}$ is the reduced Planck mass. The simplest version of dark energy is indeed that of a positive cosmological constant (CC), and so it is natural to ask if this can be incorporated within string theory. 

An alternative to the CC is the existence of a time-dependent scalar potential energy associated with a scalar field dubbed {\it Quintessence}, probably in honor of Aristotle. Although appealing, it appears highly problematic. Indeed it is plagued by two severe fine tunings: that of its energy density and that of its time variation. Why so small and why now? 

Nevertheless, the observed acceleration of the Universe is customarily associated with the value of the scalar potential at present due to an interesting coincidence between two fundamental mass scales: the scale of neutrino masses and the energy scale of dark energy. Neutrino oscillations and cosmological data give constraints on the sum of and the (squared) neutrino mass differences, $\Sigma m_\nu  \ltsim  0.1 \rm eV $
and $\Delta m_\nu^2=10^{-3}-10^{-5} \rm eV^2.$
This range of masses is remarkably close to the dark energy scale, while the origin of both quantities is very different. 

The landscape metaphor became particularly powerful in relation to the Cosmological Constant problem by offering a framework that can potentially realize anthropic selection of the cosmological constant. Indeed the multiverse equation is M = SS + EI + AP. Let us dissect this symbolic equation. 

The multiverse, M,  (pictured as a landscape) rests on three pillars: superstring  (SS) theory, which proposes many different universes, eternal inflation  (EI), which disposes (realizes) them and from which universes flow, and the Anthropic Principle  (AP), which selects and stamps the universe as good for life as we know it.  Those universes with a small positive cosmological constant allow the formation of galaxies \cite{1989RvMP...61....1W, 2018MNRAS.477.3727B}, the crucial precursor for our existence.

\subsection {Dimensional reductions}
Once awakened from the dream world of anti-de Sitter vacua, we may ask how string theory in ten dimensions, M theory in eleven and supergravity in eleven, can have any prospect of describing the four-dimensional universe that we perceive, with three dimensions of space and one of time. A possible answer is that some of the dimensions are so small and compact that they escape detection even at CERN, by high energy experiments. This sleight of hand  is known as Kaluza-Klein compactification \cite{2006hep.ph....7055R}. 

Although compact dimensions can solve the problem of unseen ones, they also give rise to a dilemma, namely, that of choosing a compactification mode (typically a Calabi-Yau manifold) among a quasi-infinity of possibilities. When compactifying string/M-theory to four dimensions, one obtains a low-energy effective theory which depends on the specific mode of folding. The number of vacua is estimated  usually to be of order $10^{500}$ and possibly to $10^{272 000}$ \cite{2015JHEP...12..164T}.

It is then wise to ask whether any compelling effective field theory coupled to gravity can be obtained as a low-energy limit of an M-theory vacuum \cite{2021arXiv211009885K}. Anyway, the landscape of possible four-dimensional low-energy effective theories arising from reduction of string/M-theory is vast. This entertains the possibility that any attractive effective field theory married to gravity can be obtained as a low-energy limit of string theory. 
However, a growing herd of swampland conjectures  suggests that this is not the case and that there is an even larger ensemble of low-energy field theories that cannot be obtained in this way.

 In particular, the AdS instability swampland conjecture asserts that nonsupersymmetric anti-de Sitter vacua are unstable and, more dramatically the (no)-dS conjecture claims, provocatively, that the de Sitter space is non-existing, thereby posing an existential threat to Einstein's cosmological constant. It has been shown that simple compactifications do not lead to any de Sitter vacua and that a highly restrictive inequality on the gradient of the 4-dimensional potential $V (\phi)$  could be derived, namely that 
$\left\lvert\Delta V\right\rvert > cV/M_{Pl}$
everywhere in field space. If true, this implies that ordinary models of early universe slow-roll inflation would be in the swampland. In terms of late-time acceleration, this precludes a positive cosmological constant, but does not forbid some form of quintessence in which acceleration is driven by a very light rolling scalar field.

\subsection {A de Sitter landscape?}
String theory opens up a {\it landscape}  of solutions (space-time vacua) with negative cosmological constant (anti-de Sitter spaces)   while most observations show that our universe undergoes an accelerated expansion, and hence is conspicuous by the presence of a positive cosmological constant, corresponding to a de Sitter space. Obtaining such de Sitter spaces remains one of the major open problems in string theory.

What seemed impossible for Vafa is not impossible for Akrami, Kalosh et al. \cite{2019ForPh..6700075A}. They start from any solution in the landscape of anti-de Sitter vacua, and transform it into a long-lived metastable de Sitter space by inserting appropriate anti-branes, or extended dynamical objects that generalize particles and strings.This mechanism gives rise to a huge landscape of string theory de Sitter vacua. Astrophysicists should not remain agnostic but need only take account of the data. If the cosmological constant imposes itself by observations, the decline of the preferred model advocated by anti-deSitter theorists will just be a faded memory. 

\subsection{The de Sitter wars}
It has proven difficult to construct de Sitter vacua in superstring theory due to the fact that the Sitter space is non-supersymmetric. Paradoxically, the dS space which is maximally symmetric does not lend itself to supersymmetry.

In practice, a de Sitter vacuum requires stabilization of all the dimension scales, or moduli, this is in general a difficult problem. Taking the technical difficulties as a hint that there is a deep resistance to constructing de Sitter vacua in string theory, Vafa and collaborators relegate de Sitter vacua to the waste basket. But the exclusion of dS solutions  from the platonic string sky \cite{2019JHEP...04..095D} is not appreciated by all members of the string tribe.
The possibility that string theory does not allow for de Sitter vacua is not new   and it would be equally dangerous to think that there is a theorem that string theory has no de Sitter vacua \cite{Woit (2018)}.  

The swampland attack  has been  countered \cite{2019ForPh..6700086K, 2018arXiv180809428K, 2019ForPh..6700075A}. Indeed the war between dS supporters and dS deniers has not ceased since the release of Susskind's article {\it The anthropic landscape of String Theory}  in 2003.  This  ignited a controversy over KKLT and the landscape, which is not even close to being extinguished. The significance of the KKLT mechanism is that it produces not one dS solution, but a wide series, thereby  feeding the credo that string theory offers an anthropic solution to the fine tuning of $\Lambda$.

The banishment of Einstein gravity and of the de Sitter space may at first appear absurd to most cosmologists since they have good evidence that the Universe is entering a phase of recent acceleration, most likely driven by a positive cosmological constant. 
However inflation is also a phase of acceleration that the Universe most likely has traversed, and it was not due to a cosmological constant. Indeed, it was most likely driven by a scalar field rolling down a potential. It is therefore not completely absurd, to consider that the late-time acceleration is also due to such a mechanism. Its motor is termed quintessence \cite{2008MPLA...23.1252M, 2020arXiv200510168V}.

The possibility that de Sitter vacua are in the swampland is therefore not ruled out by observation. It also does not mean that the anthropic selection solution to the cosmological constant problem is out of question. Fine tunings and adjustments are now concentrated on the scalar potential(s). 

\section {Marshy landscapes}
Contrary to the little warm pond dear to darwinians, the swampland of string theorists extends beyond measure, with unavoidably negative connotations. It is indeed far from established that there exists a landscape of de Sitter vacua in string theory.  On the contrary, there is mounting evidence that string theory abhors de Sitter space.  There is not a single rigorous 4D de Sitter vacuum in string theory, let alone $10^{500}$, which is quite embarrassing for the string advocates. 

There is an increasing fear that the majority, and perhaps all effective field theories,  are intolerent to gravity or {\it do not possess a sensible UV completion into a quantum theory of gravity}  in the string jargon. Such theories are  valid up to a given energy. General Relativity and the standard model of particle physics are two of them, since they cease to be relevant above the Planck energy ($\sim 10^{19}$ GeV) For the tribe of string theorists, such deficient effective theories -Quantum Field Theories in curved spacetime- are said to fall in the  {\it Swampland} of lost theories. An astute but still conjectural way of delineating the space of inconsistent theories is in the form of the famous { \it Swampland Conjectures}. 

For example, the no-global symmetry conjecture avoids overclosure of the Universe by black hole remnants of the Planck mass, the {\it distance conjecture} holds that scalar fields cannot exhibit field excursions much larger than the Planck scale, while the {\it weak gravity conjecture} states that the lightest particles of a theory cannot carry a mass larger than their charge in Planck units. The trans-planckian Swampland conjecture forbids trans-planckian quantum fluctuations from becoming classical. 

And above all, the de Sitter conjecture is especially restrictive: an issue of fundamental importance is whether effective theories that admit de Sitter vacua can be embedded within quantum gravity, or whether they sink into the Swampland. This is a topic of crucial importance since observations show that the expansion of space is accelerating under the influence of an operator resembling   Einstein's cosmological constant with an equation of state very close to $p=-D.$ 

A major objection to quintessence is that the cosmological constant problem is not resolved, since it does not prevent large contributions to the cosmological constant from quantum effects much larger that the observed dark energy density. Additionally, it exacerbates the dark energy problem by requiring that the responsible field is slowly rolling.  In short, this seems to replace a single fine-tuning by two: namely both  
$V\sim   10^{-120} M_{Pl}^4 $ and $|\Delta V| \sim
 10^{-120} M_{Pl}^3 $ need to be extremely small. 
However, this effective field theory-based reasoning may be too naive. In particular, if $|\Delta V|=cV/M_{Pl},$
 this additional fine-tuning can be avoided \cite{2020NuPhB.96015167D}. This warns us that in general the speculative swampland conjectures, are endangered by various loopholes and exceptions. 

\subsection {Swampland cosmology}
The implications of swampland conjectures have been studied in the context of cosmology, based, for instance, on the Swampland Conjecture known as the refined de Sitter conjecture which states that the effective low-energy potential $V(\Phi)$) for scalar fields $\Phi$  must satisfy
 $\Delta V >\gtsim \Delta V c M_{Pl}$ and the minimum value $\Delta^2  V \ltsim  -c^\prime/M^2_{Pl},$ 
 for universal positive constants c and c$^\prime $ of order 1, in any consistent theory of quantum gravity. If true, this would imply that the state of the universe is unstable. 

The refined de Sitter conjecture has been applied to single-field inflation models\cite{2019PhLB..788..180O}.
In particular, the ratio between scalar and tensor modes in primordial fluctuations, r, and the scalar spectral index, ns, parametrising the scale-dependence of density fluctuations, have been considered. For consistency between observational data and the model, it is found that $c ~\sim 0.1 $ is intertwined with the raw dS conjecture, and
 $ c^\prime \sim 0.01$  is in tension with the refined  version.The validity of this application has however been questioned.

\subsection {Observational tensions}
There are interesting tensions in cosmological data.  The most significant is at around a 5$\sigma$ difference between early and late epoch determinations of the Hubble constant. Early epoch refers to the distant Universe, most notably use of the temperature fluctuations in the cosmic microwave background as a distance calibrator, while late epoch refers to local distance calibrators, most notably  a laddered combination of Cepheid variable stars, the most luminous  red giant stars, and Type 1a supernovae.

Of course, similar observational issues have a long history in observational cosmology, spanning the past half century or more.  Following the triumph of Hubble in establishing the expansion of the Universe, the rate of recession of the distant galaxies remained uncertain  by some  fifty percent as Sandage and de Vaucouleurs, along with  their collaborators,  vigorously debated the distance scale throughout the 1980s. Only with the advent of large ground-based telescopes, and especially  the Hubble Space Telescope,  could one resolve Cepheid variable stars in distant galaxies and use Type Ia supernovae to establish a distance ladder to  galaxies at redshifts of 0.1 or larger.

However the distance scale controversy remains, admittedly now reduced   to roughly  ten percent uncertainty offsets  between the rival teams that apply complementary distance calibrators. This can be distilled either into a case for seeking systematic errors, or for modifying the physics of the expansion at early or late epochs.

Some authors have argued for swampland-motivated explanations \cite{2019ForPh..6700105A}  or use these tensions to set constraints on string models \cite{2019arXiv191100925L}  in order to probe dark energy \cite{2019PhRvD.100d3505C,2021PhRvD.103d3523A} or the Hubble tension \cite{2020PhRvD.101h3532A},
or even to constrain the Swampland \cite{2019arXiv190502555C, 2021PhLB..81636199A} and to favor quintessence explanations  of dark energy
\cite{2018PhLB..784..271A}, to give selected examples.

Here we simply note that Hubble constant determinations are themselves a sort of astrophysical bog where systematic and obseravational errors are uncertain. Although up to a 1\% determination of $H_0$ is claimed by Riess et al. \cite{2021arXiv211204510R}  using Cepheid-based indicators, or by other competitive methods, most notably CMB-based  \cite{2020A&A...641A...6P} and those using red giant branch calibrators \cite{2021ApJ...919...16F},  the competing results differ by far more than the quoted errors. While new physics remains one possible explanation that most notably appeals to a difference between early and late dark energy, it seems premature to take any such interpretations as robust until the observational issues are clarified and resolved.

 A decisive argument would employ high precision geometrical distance measures, such as galaxy cluster lensing-induced quasar time delays or gravitational wave sources with or without optical counterparts, but these techniques currently remain under development. For example, expected improvements in standard siren cosmology should yield better than 1\% determinations of the Hubble constant with a year of data from the Einstein telescope \cite{2021ApJ...908..215Y}.  
There are other cosmological tensions most notably from cosmic shear data, but these are currently at the 2$\sigma$ level from the Dark Energy Survey  \cite{2022PhRvD.105b3514A} and KiDS-1000  \cite{2021arXiv210904458T}.
Future surveys, especially by the Euclid,  Rubin and Roman telescopes, should significantly improve this latter constraint by a factor of two or more \cite{2021arXiv211011421N}.

\section {Comments and Conclusions}
Recent studies show that it is difficult, if not impossible, to realize de Sitter space (with a positive cosmological constant, CC) in string theory, whereas observations converge towards the existence of something like a cosmological constant playing the role of cosmic accelerator. And the balance sheet does not lean towards strings : there are no hints of extra dimensions, nor SUSY, no clear variation of physical constants, and above, there is strong empirical dominance of the despised cosmological constant, contrary to the expectations of most string aficionados.

Confronted with this situation, the more extremist string theorists practice denial of reality and throw the CC into their waste basket. The objective is to kill the CC to save a vision. But the CC is still alive. Decades of attempts to explain its tiny positive energy have left little hope for success.  

Observational prospects for establishing a fundamental challenge to $\Lambda$CDM cosmology are currently indecisive. This situation will certainly improve within a decade \cite{2021ApJ...912...99B}. For the moment however there are certainly  hints of late epoch tensions, in determinations of the Hubble constant, in the amplitude of density fluctuations, and even in the isotropy of the Universe.. But there is so far no robust evidence that favors new physics in cosmology, especially at or even before the epoch of inflation.
 
 The question of fine-tuning remains more acute than ever.  Transferring fine-tuning from the domain of the physical constants and initial cosmological conditions to the potentials of the scalar fields raises more questions than answers. 
 One remaining loophole may be to   relate the CC to the lightest neutrino mass, of meV scale, and through the see-saw mechanism to the electroweak symmetry breaking energy.  Another may lie in compacting the standard model onto a circle. 
Generous but fuzzy, our understanding of string theory is still in a state of flux. It is not yet a mature field, lacking underlying principles from which the results are rigorously derived, while  shortcomings, exceptions, counter-examples and loopholes abound. Any derivations, for instance,  are sensitive to the UV structure of the theory under test and to the swampland criteria.  
 
Nonetheless, the interest in the Swampland has not diminished, since it not only maintains vigorous discussion, controversy and hope for string theory advocates, but also,  fortunately, progressively reveals links between the different conjectures. String theorists may possibly reach a picture where there are observable consequences of an underlying new principle of order. This at least is the hope and the challenge.
 
 The swampland criteria have implications for particle physics and cosmology and astrophysics. This is why astroparticle physicists, at least by curiosity, should be aware of such speculations, taking them however with a grain of salt, and not using them in the present situation as really discriminatory. Indeed, the lack of a complete, non-perturbative definition of M-theory is a significant obstruction to any proof. Instead, the conjectures are motivated by idiosyncratic examples from string theory and black hole physics. 
 
 So far so good, but, what is more surprising to us, is that swampland advocates, on the basis of what perceive to be  fragmented lines of thought, enact definitive judgement in and on cosmology, condemning, for instance, the cosmological constant to non-existence. Cosmological data in the next decade may falsify their attempts, or even justify them,  giving information that may be not on the constraints on quantum gravity and string theory, but rather on the validity of certain swampland conjectures that exile the cosmological constant into an uninhabitable wasteland. Our preferred  conjecture is that perhaps it is just another constant of nature, to join such worthy physics compatriots as the electron mass and charge, Planck's constant, the velocity of light in vacuum and the fine-structure constant. It is inevitable that the various swampland conjectures will be modified, some ruled out by counter-examples, others refined by epicycles, and some, just conceivably, refined by future observations.

\end{document}